\documentclass[aapm,graphicx]{revtex4-1}

\draft 

\usepackage[mathlines]{lineno}

\usepackage{amssymb}
\usepackage[pdftex]{graphicx}
\usepackage{multirow}
\usepackage{color,soul}

\graphicspath{{./}}

\begin{document}
  
  \title{3D source tracking and error detection in HDR using two independent scintillator dosimetry systems}
\author{ Haydee M. Linares Rosales$^{1,2}$}
\author{Jacob G. Johansen$^{3}$}
\author{Gustavo Kertzscher$^{3}$}
\author{Kari Tanderup $^{3}$}
\author{Luc Beaulieu$^{1,2}$}
\author{Sam Beddar$^{4}$}

\affiliation{$^{1}$CHU de Quebec-Universit\'e Laval, Quebec, Canada}

\affiliation{$^{2}$ Universit\'e Laval, Quebec, Canada} 

\affiliation{$^{3}$ Aarhus University Hospital, Department of Oncology, Aarhus C, Denmark} 

\affiliation{$^{4}$The University of Texas MD Anderson Cancer Center, Houston, United States}

\email[Corresponding author: Haydee M. Linares Rosales, ]{haydee8906@gmail.com}
\date{\today}

\begin{abstract}
\scriptsize{\textbf{Purpose:} The aim of this study is to perform 3D source position reconstruction by combining in vivo dosimetry measurements from two independent detector systems.\\

\textbf{Methods}: Time resolved dosimetry was performed in a water phantom during irradiation of a $^{192}$Ir HDR brachytherapy source using two detector systems. The first was based on multiple (three) plastic scintillator detectors and the second on a single inorganic crystal (CsI:Tl). Brachytherapy treatments were simulated in water under TG-43U1 conditions, including an HDR prostate plan. Treatment needles were placed in distances covering a range of source movement of 120 mm around the detectors. The distance from each dwell position to each scintillator was determined based on the measured dose rates. The three distances given by the mPSD was recalculated to a position along the catheter ($z$) and a distance radially away from the mPSD ($xy$) for each dwell position (a circumference around the mPSD). The source x, y, and z coordinate were derived from the intersection of the mPSD's circumference with the sphere around the ISD based on the distance to this detector. We evaluated the accuracy of the source position reconstruction as a function of the distance to the source, the most likely location for detector positioning within a prostate volume,  as well as the capacity to detect positioning errors. 

\textbf{Results:} Approximately 4000 source dwell positions were tracked for eight different HDR plans. An intersection of the mPSD torus and the ISD sphere was observed in 77.2 \% of the dwell positions, assuming no uncertainty in the dose rate determined distance. This increased to 100 \% if 1$\sigma$ uncertainty bands were added. However, only 73(96) \% of the expected dwell positions were found within the intersection band for 1(2) $\sigma$ uncertainties. The agreement between the source's reconstructed and expected positions was generally within 3 mm for a range of distances to the source up to 50 mm. At distances beyond 50 mm, more significant deviations are observed. The experiments on an HDR prostate plan, showed that by having at least one of the detectors located in the middle of the prostate volume, reduces the measurement deviations considerably compared to scenarios where the detectors were located outside of the prostate volume.  The analysis showed a detection probability that, in most cases, is far from the random detection threshold.Errors of 1(2) mm can be detected in ranges of 5-25 (25-50) mm from the source, with a true detection probability rate higher than 80 \%, while the false probability rate is kept below 20 \%.  

\textbf{Conclusions:} By combining two detector responses, we enabled the determination of the absolute source coordinates. The combination of the mPSD and the ISD in vivo dosimetry constitutes a promising alternative for real-time 3D source tracking in HDR brachytherapy. 

}

\end{abstract}


\maketitle 
  
\section{Introduction}
  
    Brachytherapy (BT) is a cancer treatment modality in which radiation dose is administrated to patients through a radioactive source placed in/or close to the tumoral region. BT treatments are employed for a wide variety of cancers such as skin, gynecological, breast, prostate and lung cancer \cite{Fonseca-2017, Williamson-2006}. The high gradient field is the main advantage BT offers, allowing to deliver a high dose to the target while preserving adjacent healthy tissues and reducing the toxicity level. However, the accurate dose planning and delivery requires a very high level of precision, as a dose variation of more than 10 \% per mm can be expected close to the source. This can make even small treatment mis-administration or errors lead to nefarious consequences in patients if it goes undetected. 

    At the same time, the steep dose gradients poses a challenge for accurate dose verification. In vivo dosimetry (IVD) is the most direct measurement of the absorbed dose in regions of interest like organs at risk (OAR). IVD is therefore an ideal candidate for treatment verification during BT. Despite several studies \cite{Fonseca-2017, Cartwright-2010, Hardcastle-MOSkin-2010, Kertzscher-2011, Seymour-2011, Kertzscher-2014, Johansen-2018, Sethi-Doppler-US-2018} focusing on developing methods for real-time monitoring, nowadays there is a limited availability of commercial systems that allows for the implementation of such a technique. Thus many Centers don't perform IVD, and therefore events may remain unnoticed, or if detected they are typically only identified post-treatment \cite{Johansen-2018}. A study published by Tanderup et al. \cite{Tanderup-invivo-Brachy-2013} summarized the aspects that need to be taken into account when considering a detection system as a potential tool for IVD applications in BT. In this study, plastic and inorganic scintillators were used together. Scintillator detectors have characteristics (size, sensitivity, online read-out) that make them suitable for this kind of applications. Plastic scintillator detectors (PSDs) are furthermore water equivalent. On the other hand, they are affected by stem effect, which is the contaminating Cherenkov and fluorescence light induced in the fiber-optic cable. Stem effect can cause large deviations in the measured dose rate for/in/during HDR BT with an $^{192}Ir$ source, if not taken into account \cite{Therriault-2011, Therriault-Phantom-2011}. The production and removal of stem effect in PSDs are widely discussed topics \cite{Beddar-water-equivalent-1992-1, Beddar-water-equivalent-1992-2, Boer-optical-1993, Fontbonne-2002, Clift-temporal-2002, Lambert-Cerenkov-2008, Archambault-MathForm-2012, Beaulieu-Review-2016}. Furthermore, the use of PSD in a multipoint configuration (mPSD) could assess the dose at multiple points simultaneously, thereby improving treatment verification quality and accuracy \cite{Therriault-mPSD-2012, Therriault-mPSD-Brachy-2013, Patricia-2016}. The Inorganic scintillator detectors (ISDs) exhibit light yields (photons/keV) 2-3 orders of magnitude greater than PSDs  \cite{Kertzscher-Inorganic-Scint-2017}. The high signal intensity reduces the impact of the stem signal to a neglible level. Furthermore, it enable the use of smaller scintillator volumes that measure dose rates with small statistical uncertainties. ISDs are however not water equivalent and they exhibit an energy dependence on the photon spectrum they are exposed to. 
    
    This paper will focus on the use of four scintillator point detectors for  real-time 3D source tracking in HDR BT. The study was performed with an ISD  and a mPSDs systems. The latter containing three PSDs as described by Linares et al. \cite{Linares-2019-dosimetry-mPSD}. Both mPSD and ISD have previously been used for source tracking during BT. In both cases a degeneracy in the azimuthal angle was observed \cite{Linares-2019-dosimetry-mPSD, Johansen-2018}. In this paper, the two systems are combined to resolve this degeneracy.

\section{Materials and Methods}
    
    \subsection{The detector systems}

    Two independent detector systems were used in this study, one for the mPSD and one for the ISD. These are described in this section.

    \subsubsection{mPSD system}
    
    One of the systems used for HDR BT was a 1 mm diameter mPSD composed by organic scintillators BCF-10, BCF-12 and BCF-60 scintillators from Saint Gobain Crystals (Hiram, Ohio, USA), with lengths of 3 mm, 6 mm, and 7 mm respectively. The scintillators and optical fiber were tightly light-shielded. The mPSD was coupled to 15 m long optical fibre type \textit{Eska GH-4001} from Mitsubishi Rayon Co., Ltd. (Tokyo, Japan) connecting to a light collection box. The box consisted of a beam aligner block A10760 from Hamamatsu (Bridgewater, USA) \cite{PMT-Hamamatsu} coupled to an Olympus infinity-corrected objective lens RMS40X from Thorlabs (Newton, USA); as well as photomultipliers tubes (PMT) coupled to a set of dichroic mirrors and filters. Four sets of PMTs, mirrors and filters were used, one for each scintillator and one to account for the stem effect \cite{Therriault-2011, Archambault-MathForm-2012}. A detailed description of the entire system can be found in \cite{Linares-2019, Linares-2019-dosimetry-mPSD}.The signal was read and sent to a computer at a rate of 100 kHz using a data acquisition board (DAQ) type DAQ NI USB-6216 M Series Multifunction from National Instruments (Austin, USA) \cite{DAQ-6289}. The scintillation light detection system is independently controlled from the irradiation unit with homemade software based on Python.

    \subsubsection{ISD system}
    
    The ISD consisted of a 1.0 mm-diameter and 0.5 mm-long scintillating crystal made of CsI:Tl that was coupled to a 1 mm-diameter and 15 m-long fiber-optic cable. The crystal and optical fiber were shielded from light using an opaque plastic tube. A more detailed characterization of the CsI:Tl crystal can be found in \cite{Kertzscher-Inorganic-Scint-2017, Kertzscher-Inorganic-scint-2019}. The optical fiber was connected to a Si-diode photodetector (s8745-01 from Hamatsu), which was connected to a data acquisition system (usb2408 from Measurement Computing) that monitored the signal with a 30 kHz sample rate. The signal was averaged across 50 ms internally in the data acquistion system before sent to the user computer at a rate of 20 Hz.

    \subsection{Dose measurements}
    
    Dose measurements were carried out in a 40 x 40 x 40 $cm^{3}$ water tank with an additional layer of 10 cm solid water (Plastic Water Standard, CIRS, Virginia, USA) on each side to ensure TG41-U1 conditions \cite{TG-43-Update}. A Flexitron HDR afterloader from Elekta (Elekta Brachy, The Netherlands) coupled to 30 cm needles from Best Medical International (Springfield, VA, USA), was used for irradiation. The needles were placed in the water tank in a series of different configurations (see section \ref{MM_study_cases}) by means of a custom-made poly(methyl methacrylate) holder composed of two needles insertion templates of 12 x 12 $cm^2$, each having 225 holes in a 5 mm spacing  grid\cite{Therriault-mPSD-Brachy-2013, Linares-2019}. Both template were placed 20 cm apart . All needle ends were placed at the same height. The mPSD and ISD dimensions allowed them to be inserted into an additional needle each. The guidetubes were connected to the treatment needles, before the two detectors were placed in their respective needles. The dose rate was recorded with both systems and analysed post-irradiation.
    
    \subsubsection{Study cases}
    \label{MM_study_cases}
    
\begin{figure}[h!]
    \centering
    \setlength{\tabcolsep}{-15pt}
    \begin{tabular}{ccc}
    &\includegraphics[trim = 0mm 0mm 0mm 0mm, clip, scale=0.45]{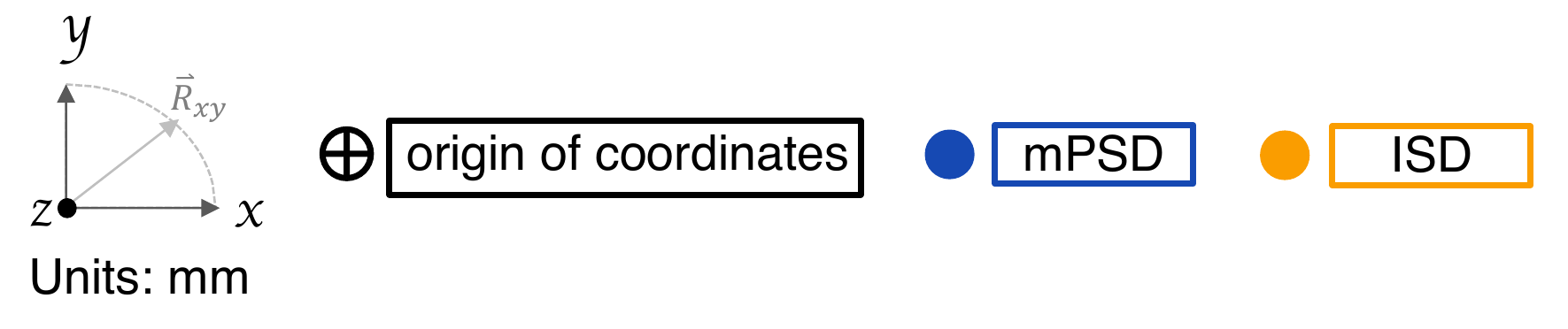}\\
    \includegraphics[scale=0.43]{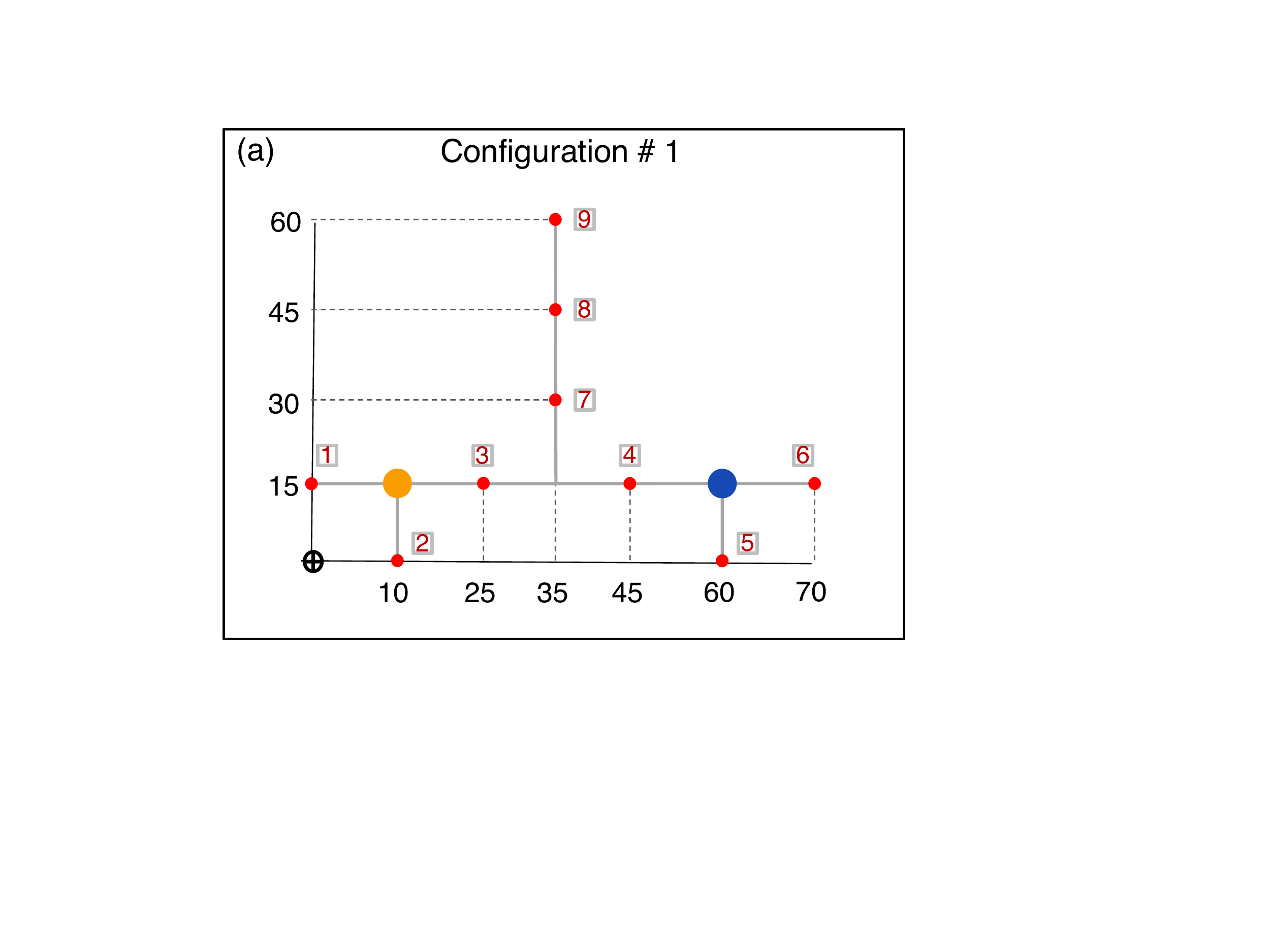}&
    \includegraphics[scale=0.43]{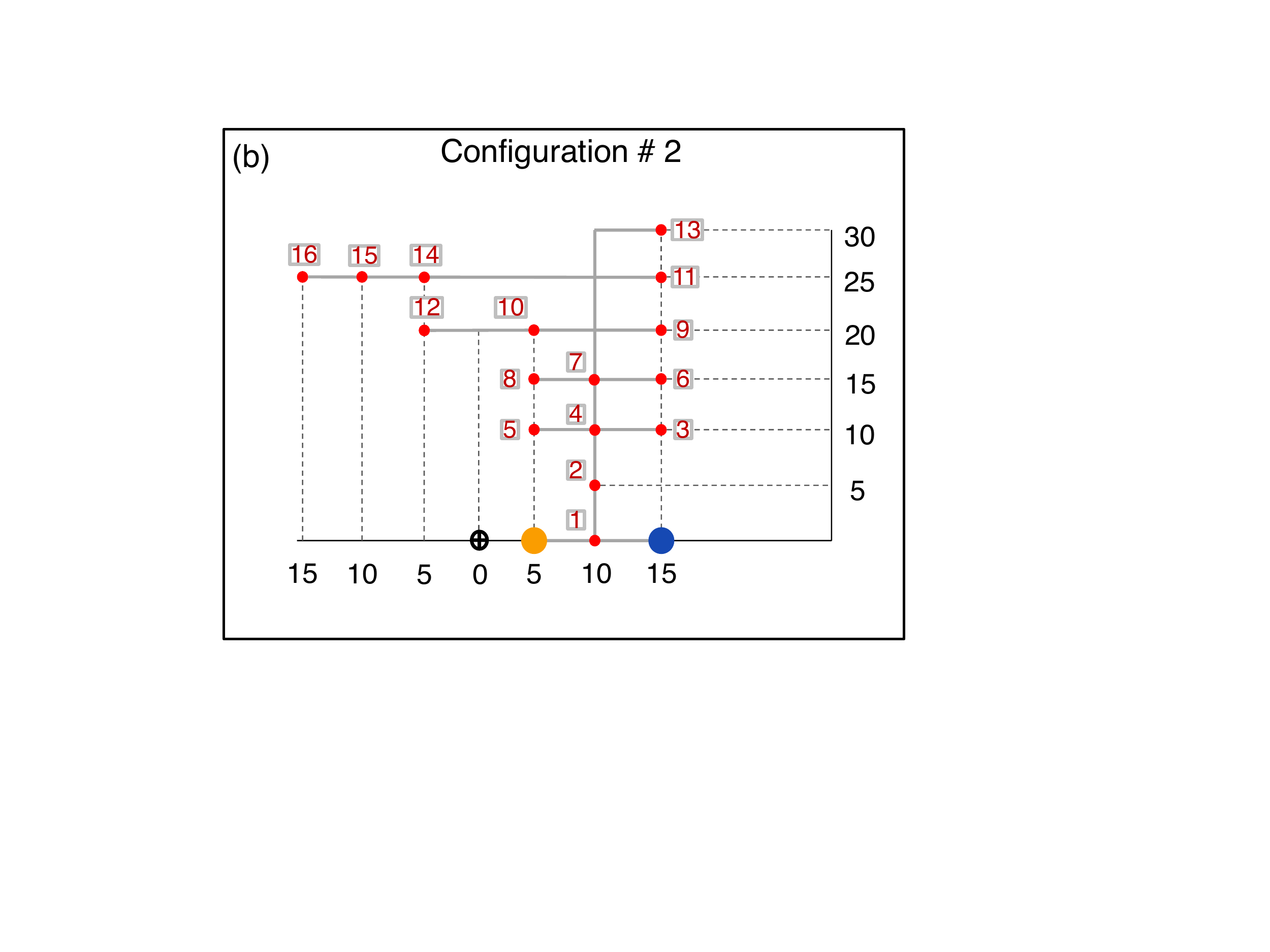}&
    \includegraphics[scale=0.43]{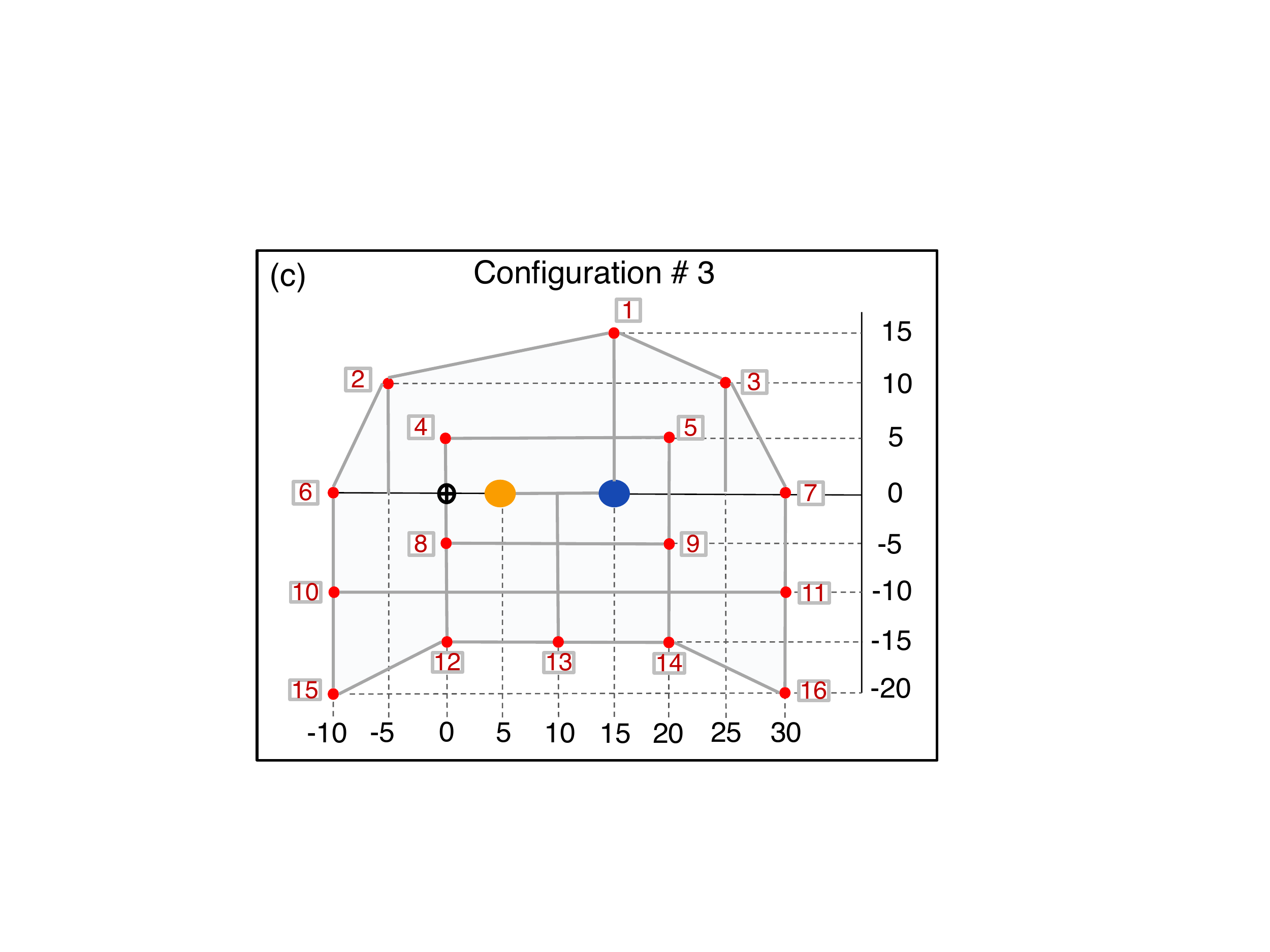}
    \end{tabular}
    \caption{\label{simult_dose_meas_scheme} Schematic of the 2 detectors and needles distribution in the x-y plane during in-water measurements with the mPSD and ISD systems. Measurement set-up when studying the dosimeters response in a symmetric (a) and asymmetric (b) configurations. (c) Needle positions during the HDR prostate plan irradiation. The detectors are placed as in plan 5.}
    \end{figure}

    Three different treatment configurations were used as study cases. Figure \ref{simult_dose_meas_scheme} shows a schematic of the two detectors and needles distribution in the x-y plane.  The first type of treatment was used to study the response of the dosimeters when arranged in a symmetric configuration relative to the detectors as shown in figure \ref{simult_dose_meas_scheme}(a). The detectors were surrounded by nine hollow plastic needles at different distances in the x-y axis. In the second treatment configuration (figure \ref{simult_dose_meas_scheme}(b)) sixteen needles were positioned asymmetrically around the detectors in a very confined space. The third type of treatment used in this study was an HDR prostate plan. The needles, detectors, as well as the custom-made poly(methyl methacrylate) holder were scanned using a CT on rails Somaton Sensation Open from Siemens (Malvern, USA). The CT scan geometry was exported to Oncentra Brachy v4.0 (Elekta Brachy, The Netherlands) where the plan was created with a prescribed dose per fraction of 15 Gy. Figure \ref{simult_dose_meas_scheme}(c) shows a schematic of the template used for a prostate treatment irradiation. Sixteen needles were reconstructed and used for treatment delivery. Several measurements were done in order to evaluate the benefit of positioning the mPSD and ISD in different locations inside the treatment volume. 
    
    Table \ref{table_MM_Irr_plan_info} details the characteristics of the plans created. The first column in table \ref{table_MM_Irr_plan_info} shows the set-up configuration used for measurements according to figure \ref{simult_dose_meas_scheme}. The second column refers to the plan number assigned to each test and will be used in the further sections. The third and fourth columns in table \ref{table_MM_Irr_plan_info} show the x,y,z coordinates of the position of the ISD and mPSD respectively. The mPSD z-coordinate refers to BCF10's position. The fifth column in table \ref{table_MM_Irr_plan_info} shows the number of needles used for irradiation. The sixth column shows the number of dwell positions planned at each needle, while in column seven are shown the lowest/highest distances explored in the z-axis. Column number eight contains the dwell position steps planned for each treatment. The last column in table \ref{table_MM_Irr_plan_info} refers to the dwell times used for measurements. 
    
    \begin{table}[ht]
    \caption{\label{table_MM_Irr_plan_info} Irradiation plan parameters used to study the mPSD and ISD systems response. The number of dwell position, source step and dwell time varied from position to position in configuration 3. Therefore only the ranges [min-max] are given.}
    \setlength{\tabcolsep}{5pt}
    \centering
    \resizebox{\textwidth}{!}{
    \vspace{0.2cm}
    	\begin{tabular}{ccccccccc}
    	\hline
    	\hline
    	\multirow{2}{*}{\textbf{Configuration}}
    	& \multirow{2}{*}{\textbf{Plan}} 
    	& \textbf{ISD } 
    	& \textbf{mPSD} 
    	& \textbf{Number} 
    	& \textbf{Number of}
    	& \textbf{lowest/highest} 
    	&\textbf{Source}   
    	& \multirow{2}{*}{\textbf{dwell t}} \\
    	                       
    	&              
    	& \textbf{position} 
    	& \textbf{position} 
    	& \textbf{of needles}
    	& \textbf{dwell p.}
    	& \textbf{Z} 
    	&\textbf{step}   
    	& \\
    	                               
    	& 
    	& \textbf{(mm)} 
    	& \textbf{(mm)} 
    	&     
    	&         
    	& \textbf{(mm)} 
    	& \textbf{(mm)} 
    	& \textbf{(s)} \\
    	\hline
        \multirow{1}{*}{\textbf{1}} 
                                           & 1 & (10, 15, 0) & (60, 15, -4.4)  & 9   & 21  & -31/+69  & 5  & 5  \\
    	\hline
    	\textbf{2}                   & 2 & (5, 0, 0)   & (15, 0, 19)   & 16  & 101  & -25.8/+74.2   & 1  & 3 \\ 
    	\hline
    	\multirow{5}{*}{\textbf{3}} &	3  & (5, 0, 0)   & (15, 0, -6.8) & 16  & 2-8 & +18.6/+53.6 &  5-30 & 0.2-34.1    \\ 
    	& 4             & (10, -20, 0)                  & (15,  0,  -4.2)  & 16  & 2-8 & +20.6/+55.6 & 5-30 & 0.2-34.1  \\
    	& 5             & (10,  20, 0)                  & (15,  0,  -6.0)  & 16  & 2-8 & +19.2/+54.2 & 5-30 & 0.2-34.1  \\
    	& 6             & (10,  20, 0)                  & (10, -20, -4.0)  & 16  & 2-8 & +17.6/+52.6 & 5-30 & 0.2-34.1  \\
    	& 7             & (0, -20,  0)                  & (20, -20, -4.4)  & 16  & 2-8 & +18.2/+53.2 & 5-30 & 0.2-34.1 \\
    	& 8            & (5, 0, 0)                     & (15, 0, -6.8)    & 16  & 121 & -44/+76 & 1 & 1 \\
        \hline
    	\end{tabular}}
    \end{table}
    
    \subsection{Calibration}
    
    \label{MM_sect_calibration}
    
    A calibration is needed to transform the recorded signal into dose rate for each system. The calibration for the mPSD included removal of the stem signal, while the calibration for the ISD included an energy correction term. Both calibrations were carried out using measurements obtained under the same experimental conditions as described above.
    
    The calibration matrix and dose values for the mPSD were calculated according to the formulation published by Linares et \textit{al.} \cite{Linares-2019} for a 3 points mPSD configuration, based on the formalism proposed by Archambault et al. \cite{Archambault-MathForm-2012}. Thus the quantity of interest, dose, is insensitive to stem effect contribution.
    
    The ISD calibration was performed using a series of dwell positions in needle 1 of configuration 1. A total of 101 dwell positions were used with dwell step size of 1 mm and dwell times of 5 s. The shape of the measured dose rates were used to determine the point in the source needle closest to the ISD. The actual distance to the detector for each dwell point could then be derived. A calibration factor (CF) was determined as the ratio of the measured signal and the dose rate based on TG43 for each of the points taking into account the energy correction, which was determined prior to the calibration. The energy dependency was determined by measuring the signal from the source at a range of source positions. The positions covered distances from 5 mm to 50 mm and angles from 10 degrees to 170 degrees. The ratio of the measured signal and the dose rate based on TG43 was determined at each point and normalised at 10 mm, 90 degree. The ratio was then plotted as function of the measured dose and fitted to polynomial function. The equation for transforming measured signal (MS) from the ISD to dose rate (DR) is given in equation \ref{MM_eq_DR_CsI}.
    \begin{equation}
    \label{MM_eq_DR_CsI}
        DR = \frac{MS \cdot CF}{3.872\cdot10^{-6} \cdot (MS/S_k)^{-0.8083}+0.6517}
    \end{equation}
    Here $S_k$ is the source strength.
     
    \subsection{3D source position reconstruction}
    
    \subsubsection{From dose rate to distance to the source}
    \label{MM_sec_r_calculation}
    
    The dose rate for a given dwell position was determined for each of the four sensitive volumes by averaging the signal in Volt recorded during the dwell. This average value was then converted into dose rate using the calibration described in section \ref{MM_sect_calibration}. The distances between the individual source positions and the individual sensitive volumes were determined based on the relation between dose rate and distance on TG43-U1 formalism $r(\theta)_{i,j} = f_{TG43}^{-1}(DR)$. Here $i$, $j$ is the source dwell position and the four active volumes/crystals respectively. $f_{TG43}^{-1}$ is the inverse function of the TG43-U1 formalism. The angular dependency stems from the anisotropy of the dose distribution. In the same manner, the uncertainty of the mean dose rate were determined and transformed into an uncertainty in the distance $\sigma_{r_{i, j}}$. Thus, a distance and uncertainty between source and sensitive volume were determined for each dwell position and each sensitive volume.
    
    The relative position of the four sensitive volumes are known as well as the distance from a single dwell position to each volume. This enables a determination of the most probable dwell position based on dose rate, as described in \ref{MM_sec_3D}.
    
    \subsubsection{Needle offset calculation}
    \label{MM_needle_shift}
    
    The positioning and CT reconstruction of the needle and dosimeters lead to a positional uncertainty in the order of 0.5 mm. To reduce this positional uncertainty, needles offset were calculated using the radial distances measured with the mPSD's sensors and the ISD. According to the recommendations provided by Johansen et \textit{al.} \cite{Johansen-2018}, the distance offset of the measurements, was separated into a shift along the needle ($z$) and one radially away from the dosimeter ($r$). This was done by making a joint virtual shift of all the dwell positions in a single needle to best match the measured dose rates to TG43. The needle shift was found applying the \textit{Dynamic Time Warping} (DTW) method. DTW is a well-known technique to find an optimal alignment between two given sequences. It is a point-to-point matching method under some boundary and temporal consistency constraints. The sequences are warped in a nonlinear fashion to match each other \cite{Rabiner-DTW-1993}.  
    
\begin{figure}[h!]
    \centering
    \begin{tabular}{c}
    \includegraphics[trim = 0mm 0mm 0mm 0mm, clip, scale=0.63]{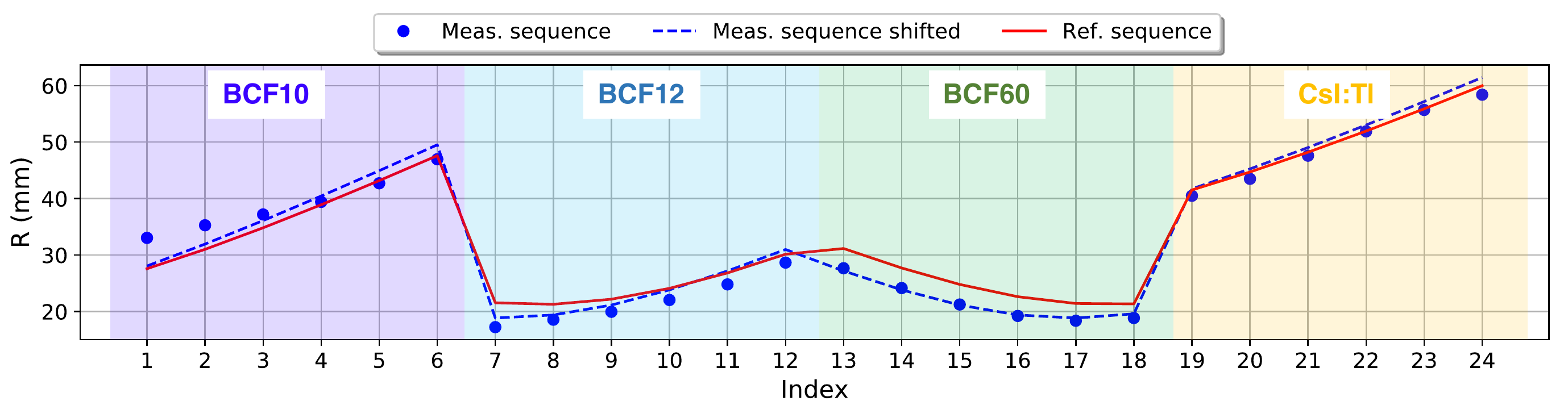}
    \end{tabular}
    \caption{\label{MM_combined_shift_calc} Example of the needle shift fit routine. The blue dots are the measured radial distances (measurement sequence) for the dwell positions, the red line is the expected curve based on the planned positions (reference sequence), the dashed-blue curve is after the shift in needle position. The inserted shaded regions represent the index limits for the data collected by each scintillator. For interpretation of the references to color in this figure legend, the reader is referred to the Web version of this article.} 
    \end{figure}

    Figure \ref{MM_combined_shift_calc} shows an example of the needle shift fitting routine used in this paper. The needle shifts calculations were done following 2 steps. We first built the 2D-shaped sequences of data to be used. The first sequence was created using the ensemble of radial distances measured by the four detectors (measurement sequence, blue dots in figure \ref{MM_combined_shift_calc}). The second sequence was used as a reference and built based on the planned radial distance to each one of the detectors (red line in figure \ref{MM_combined_shift_calc}). Both, the measurement and reference sequences were created as a function of the afterloader indexer or index number, which is the shared information between the detectors. However, since the shift analysis involved four scintillators, for each needle, the index number's array had dimensions equal to four times the afterloader indexer. Thus, the sequences' information was sorted concatenating the information for each one of the scintillators in the following order: BCF10, BCF12, BCF60, CsI:Tl. In a second step, our DTW algorithm simulated shifts in the reference sequence's $r$ and $z$ directions to find the optimal fit to the measured data (dashed-blue lines in figure \ref{MM_combined_shift_calc}). Since at short distances to the source miss-positioning errors has more remarkable effects and the measurement uncertainty increases with the distance to the source \cite{Andersen-time-resolved-2009}, the inverse square of the distance to the source was used as a weighting factor for DTW calculations. A ($z,r$) offset were found for all needles in all the configurations in Table \ref{table_MM_Irr_plan_info}, and the updated source positions were used as the ground truth in the further analysis.
    
    \subsubsection{3D source position reconstruction}
    \label{MM_sec_3D}
    
\begin{figure}[h!]
    \setlength{\tabcolsep}{1pt}
    \centering
    \begin{tabular}{cc}
    \includegraphics[trim = 1mm 1mm 1mm 1mm, clip, scale=0.36]{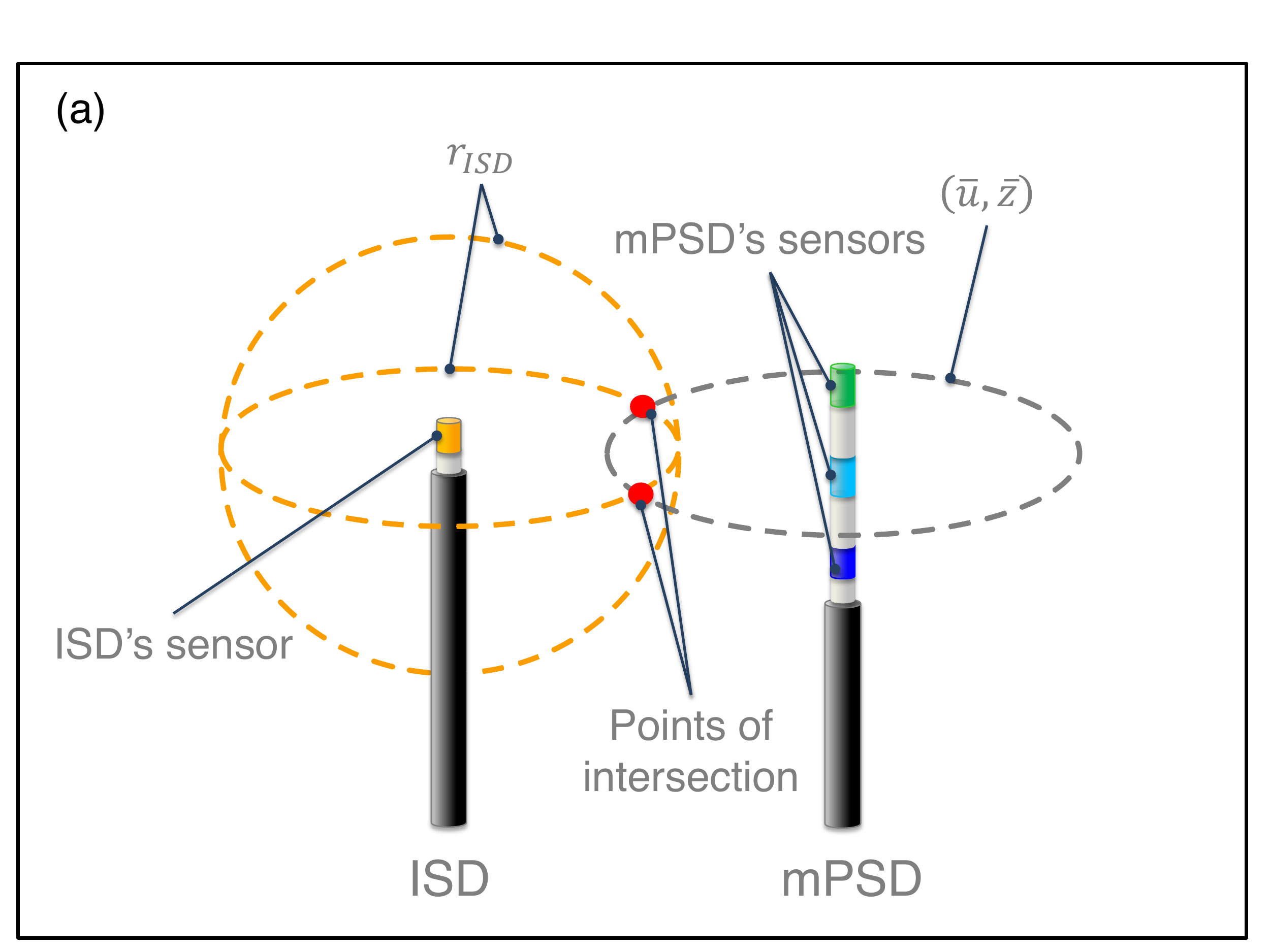}&
    \includegraphics[trim = 1mm 1mm 1mm 1mm, clip, scale=0.36]{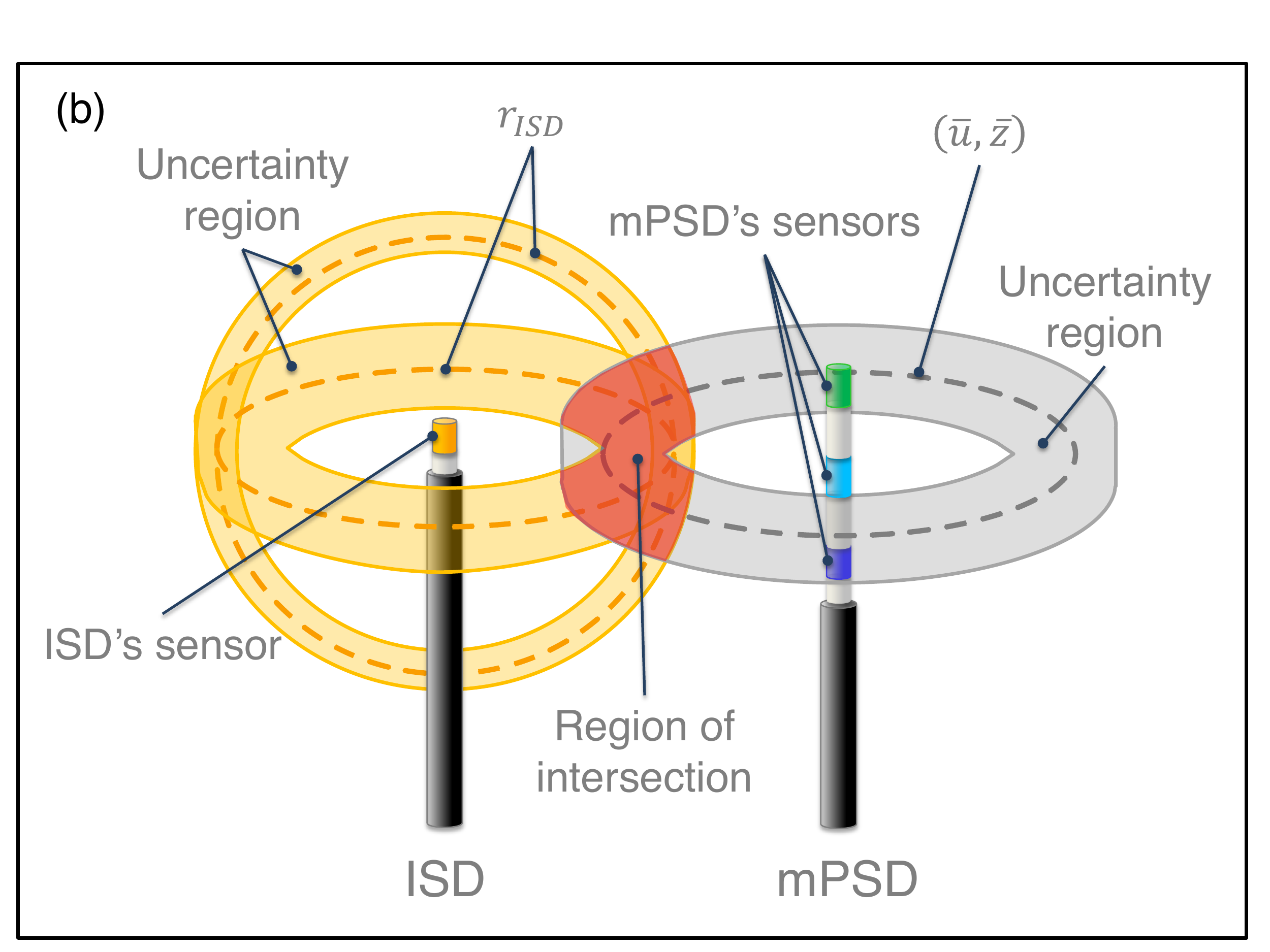}\\
    \multicolumn{2}{c}{ \textbf{\includegraphics[trim = 1mm 1mm 1mm 1mm, clip, scale=0.36]{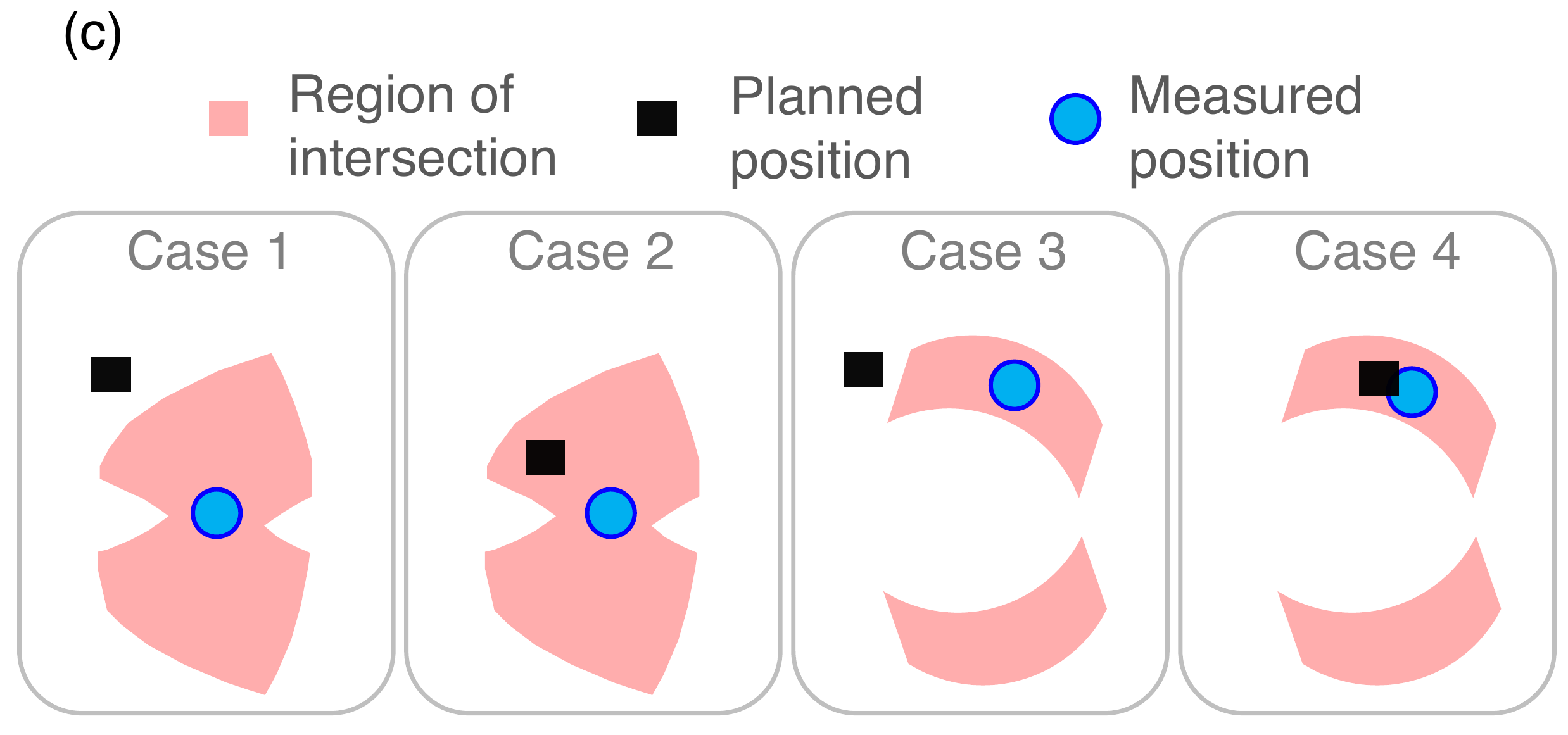}}}
    \end{tabular}
    \caption{\label{intersection_scheme_CsI_mPSD} Schematic that illustrates the combination of mPSD-ISD responses to determine the source position using no uncertainties (a) and non-zero uncertainties (b), and possible cases (c) that can arise when extracting the intersection region in (b). }
    \end{figure}

     The 3D $^{192}Ir$ source location in the space was calculated finding the intersection between the circumference predicted with the mPSD and the sphere around the single ISD. Figure \ref{intersection_scheme_CsI_mPSD} shows the schematic of the mPSD-ISD responses (degeneracy) to determine the source position in a 3D space. Figure \ref{intersection_scheme_CsI_mPSD}(a) shows the schematic when the measurement uncertainty is excluded from the source location calculation. In such a case, and based on the distance to the detectors, the goal is to find the intersection points between the mPSD's predicted circle and the sphere around the ISD. When the mPSD-ISD intersected once, the intersection point was used as source location. In cases where two intersection points were found, the closest point to the planned location was used as source coordinate. In case of no interception, we selected, for the mPSD, the point on the torus (figure \ref{intersection_scheme_CsI_mPSD}) closest to the planned position, and for the ISD, the point on the sphere closest to the planned position. Then the average of these two coordinates was used as reconstructed source position.
    
    Figure \ref{intersection_scheme_CsI_mPSD}(b) shows a schematic of the intersection areas obtained when including the uncertainty budget of each dosimeter in the calculations. The uncertainty in the measured dose rates were transformed into uncertainty bands on the distances, and the intersection areas were determined rather than intersection points. In figure \ref{intersection_scheme_CsI_mPSD}(c) are represented the possible cases that can arise when extracting the intersection region in \ref{intersection_scheme_CsI_mPSD}(b). The source location was defined as the mean coordinate defined by the region. In cases where more than one intersection band was found (figure \ref{intersection_scheme_CsI_mPSD}(c)'s case 3 and 4), the closest band to the planned source position was selected.
    
    \subsection{Detection of positioning errors}
    
    The last experiment performed in this study was to evaluate the capacity of the mPSD-ISD combined system to detect treatment positioning errors in HDR BT with an $^{192}Ir$ source. This analysis was done in two stages. We initially derived the  Receiver Operating Characteristic (ROC) curve and based on it, we analyzed the performance through positioning errors introduced in configuration \#1 and \#3 in figure \ref{simult_dose_meas_scheme}. The reasoning to build the ROC, was to establish a metric that allows during real-time acquisitions the classification of deviations into possible errors with certain level of confidence. 
    
    \begin{figure}[h!]
    \centering
    \begin{tabular}{c}
    \includegraphics[trim = 0mm 0mm 0mm 0mm, clip, scale=0.55]{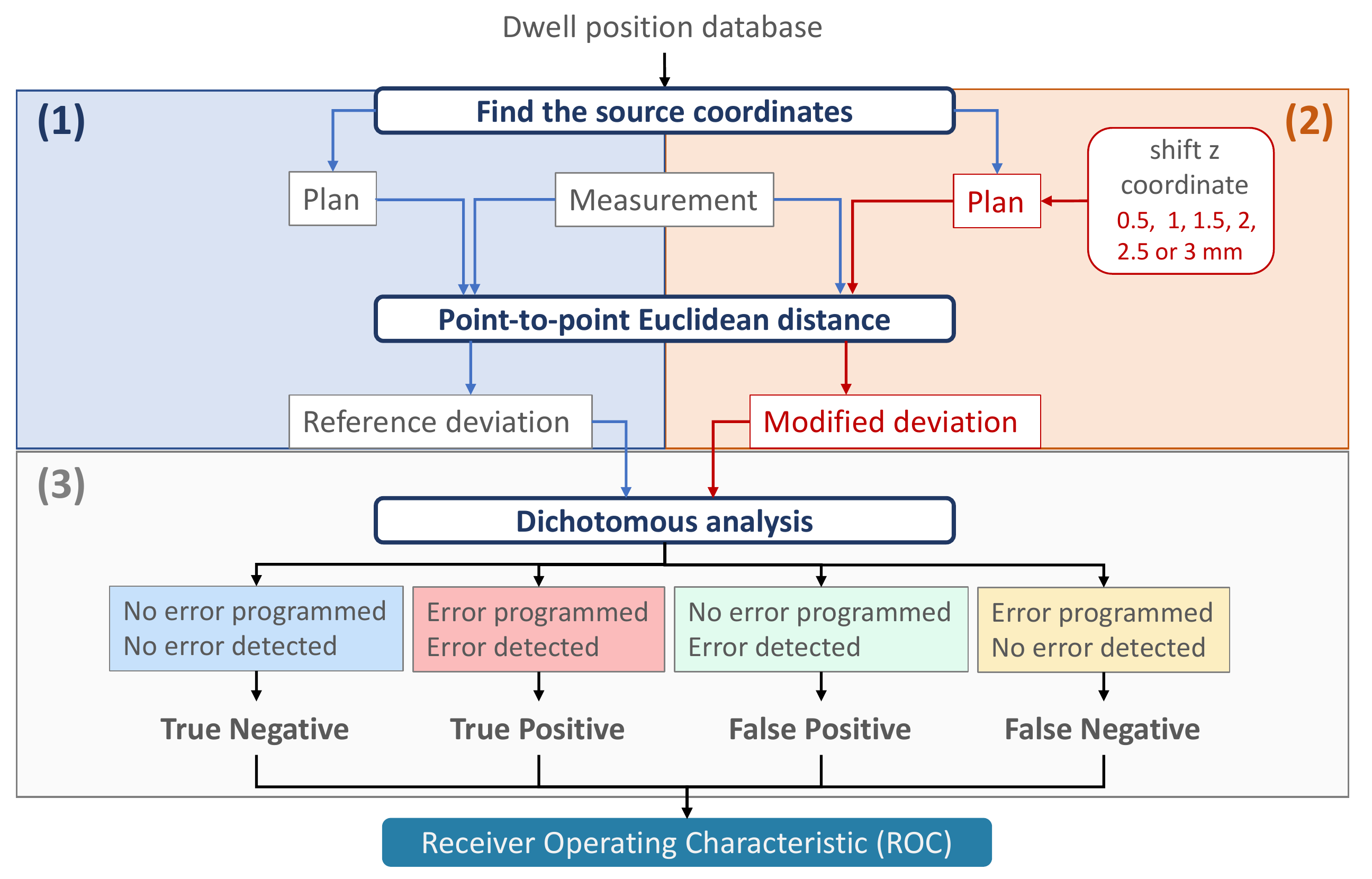}
    \end{tabular}
    \caption{\label{MM_ROC_workflow} ROC calculation workflow.} 
    \end{figure}
    
    Figure \ref{MM_ROC_workflow} shows the workflow followed to obtain the ROC curve. We build a database of information with 3741 dwell positions from plans number 1, 2, and 8 in table \ref{table_MM_Irr_plan_info}. Three main steps were followed. In step 1 we extracted the planned and measured source coordinates in a 3D space after offset correction as described in section \ref{MM_needle_shift}. We then calculated the point-to-point \textit{Euclidean distance} between the measured and planned location. This array of \textit{Euclidean distances} is referred as the \textit{reference deviation}. In step 2, we used the measured source coordinates without modifications. However the planned coordinates were modified to simulate positioning errors.  Dwell positions were randomly selected, and errors of 0.5, 1.0, 1.5, 2.0, 2.5 and 3.0 mm were introduced by shifting the $z$ coordinate. The point-to-point \textit{Euclidean distance} was then calculated using the measured and the ``modified'' source coordinates.  This data-set of deviations is referred as \textit{modified deviations}. The dichotomous analysis is performed in step 3. We used  the \textit{reference} and \textit{modified} deviations to quantify the sensitivity/specificity of the dosimeters in detecting the errors introduced. For each error simulated, a contingency table was built associating an error to a deviation beyond a given range. Those dwells where no error was programmed and the system detected no error constituted the true negatives, while the dwells where an error was programmed and detected constituted the true positives. If at a specific location, the system detected a condition different from the reference, then we classified that deviation as false-negatives or false-positives. Since the error detection is influenced by the measurement uncertainty \cite{Andersen-time-resolved-2009, Tanderup-invivo-Brachy-2013}, the ROC analysis was done for three ranges of radial distances with respect to the source independently: 5 - 25 mm, 25 - 50 mm and 50 - 75 mm. 
    
    Different plans were irradiated with and without modifications made to the reference source position. We aimed to simulate different types of dose delivery errors that current afterloader safety systems are unable to detect. The study was conducted using the configuration number 1 and 3 in table \ref{table_MM_Irr_plan_info}. In configuration number 1, the needles 5 and 6 were exchanged; and wrong dwell positions were planned in needles 3, 7, and 9. For the irradiations in a prostate template (configuration number 3), plan 3 constituted the reference. Needle 5 and 9 were swapped, as well as needle 12 with 13, and 15 with 16.  The deviations were classified into error or not, using the ROC curves as the metric and a \textit{distance-based threshold selection}. In the \textit{distance-based threshold selection} approach, the threshold to consider a deviation as an error was set by fixing the combination of false positive rate (FPR) at 20 \% and the true positive rate (TPR) at 80 \%. Thus for dwell positions with a planned distance to the source within 5 to 25 mm range, the deviation threshold used was 1 mm; for dwells falling within 25 to 50 mm range, the deviation threshold used was of 2 mm; and for the dwell positions with distances to the source beyond 50 mm, the deviation threshold used was 3 mm.

\section{Results}

    \subsection{Measurement uncertainty influence on the source position reconstruction accuracy}
    
\begin{figure}[h!]
    \centering
    \begin{tabular}{c}
    \includegraphics[trim = 0mm 0mm 0mm 0mm, clip, scale=0.4]{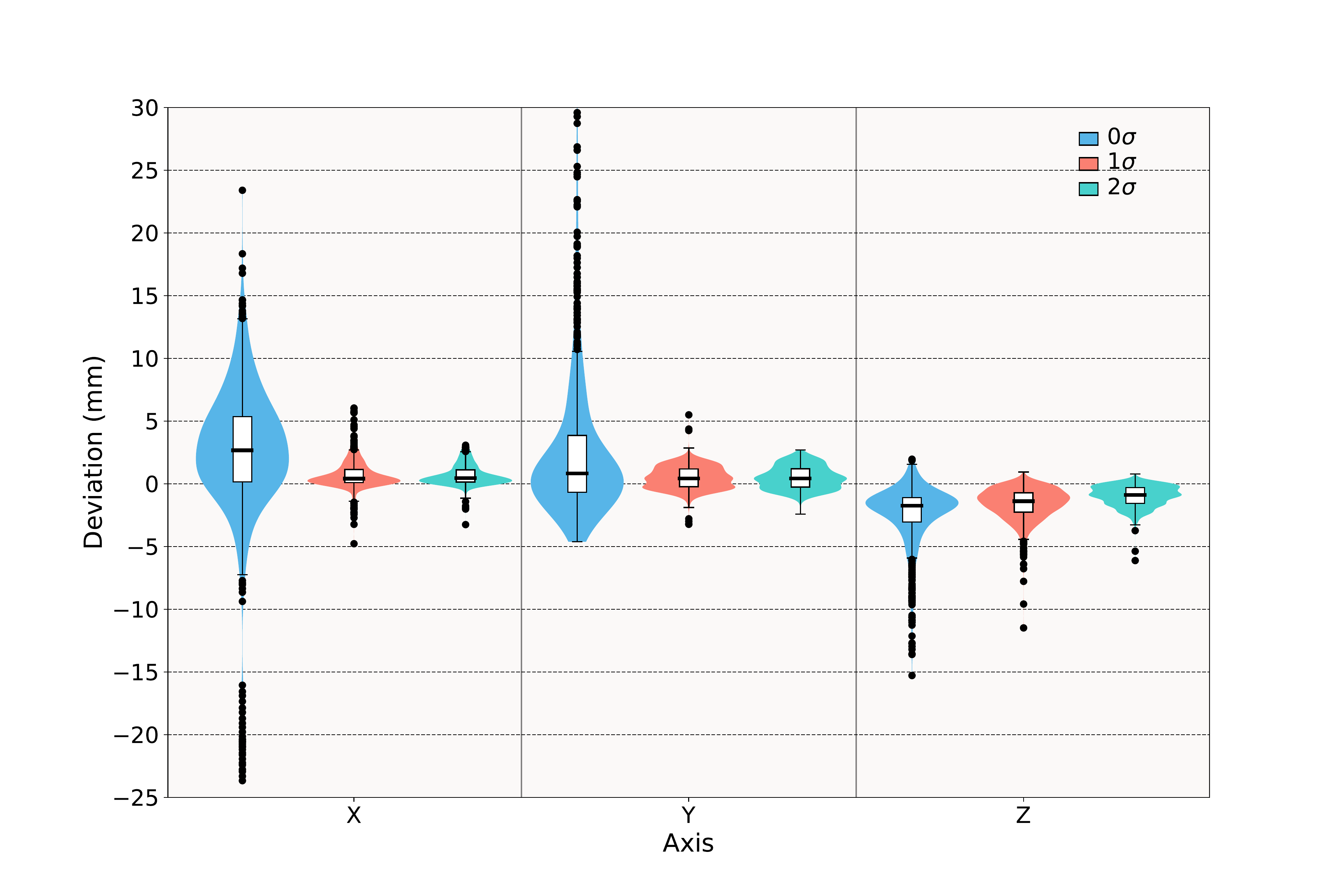}
    \end{tabular}
    \caption{\label{R_mPSD_CsI_violin_plot} Density distribution of deviations between the planned and measured position. Boxes represent interquartile ranges (quartile 1 and quartile 3), and the solid line inside the boxes represent the median values. The inner vertical lines extend from each quartile to the minimum or maximum.} 
    \end{figure}

    A total of 3741 source dwell positions were tracked while using the plans number 1, 2, and 8 in table \ref{table_MM_Irr_plan_info}. A range of 120 mm in the $z$-axis was covered. The violin plots \cite{Hintze-Violin-Plot-1998} in figure \ref{R_mPSD_CsI_violin_plot} show the density distributions of the differences between the measured source position and the planned position in the $x$, $y$ and $z$ axis. The inner boxes represent the interquartile ranges, and the line inside the boxes indicate the median values of the distributions. The black dots in each deviation distribution represents the outliers (points which falls more than 1.5 times the interquartile range above the third quartile or below the first quartile). The analysis performed assuming no measurements uncertainty in extracting the source location calculation is represented by $0 \sigma$ in figure \ref{R_mPSD_CsI_violin_plot}. $1 \sigma$ and $2 \sigma$ in figure \ref{R_mPSD_CsI_violin_plot} refers to the deviations obtained when the uncertainty in the measured dose rates were transformed into uncertainty bands to find the source location. Table \ref{table_R_tracking_uncertainty} summarizes the influence of the uncertainty budget in the measurement accuracy of the source location in a 3D space. Number of dwells with intersections refer to those dwell positions where an intersection between the mPSD's predicted ring and the sphere predicted by the ISD intersect. The analysis shown in the following sections corresponds to a source tracking with 2$\sigma$ bands criterion.
    
    \begin{table}[h!]
    \caption{\label{table_R_tracking_uncertainty} Influence of the uncertainty budget in the measurement accuracy of the source location in a 3D space.}
    \setlength{\tabcolsep}{8pt}
    \centering
    \vspace{0.2cm}
    	\begin{tabular}{cccc}
    	\hline
    	\hline
    	& \textbf{0 $\sigma$} & \textbf{1 $\sigma$} & \textbf{2 $\sigma$} \\
    	\hline
    	\textbf{Number of dwells}  & \multirow{2}{*}{77.2 \%}	& \multirow{2}{*}{100 \%}	& \multirow{2}{*}{100 \%} \\
    	\textbf{with intersection}   \\
    	\textbf{Expected positions within} & \multirow{2}{*}{Not applicable}	& \multirow{2}{*}{73 \%}	 & \multirow{2}{*}{96 \%} \\
    	\textbf{the intersection area}  \\
    	\hline
    	\end{tabular}
    \end{table}
    
    \subsection{Source position reconstruction accuracy as function of the  distance to the source}
    
\begin{figure}[h!]
    \setlength{\tabcolsep}{1pt}
    \centering
    \begin{tabular}{cc}
    \includegraphics[trim = 1mm 1mm 1mm 1mm, clip, scale=0.20]{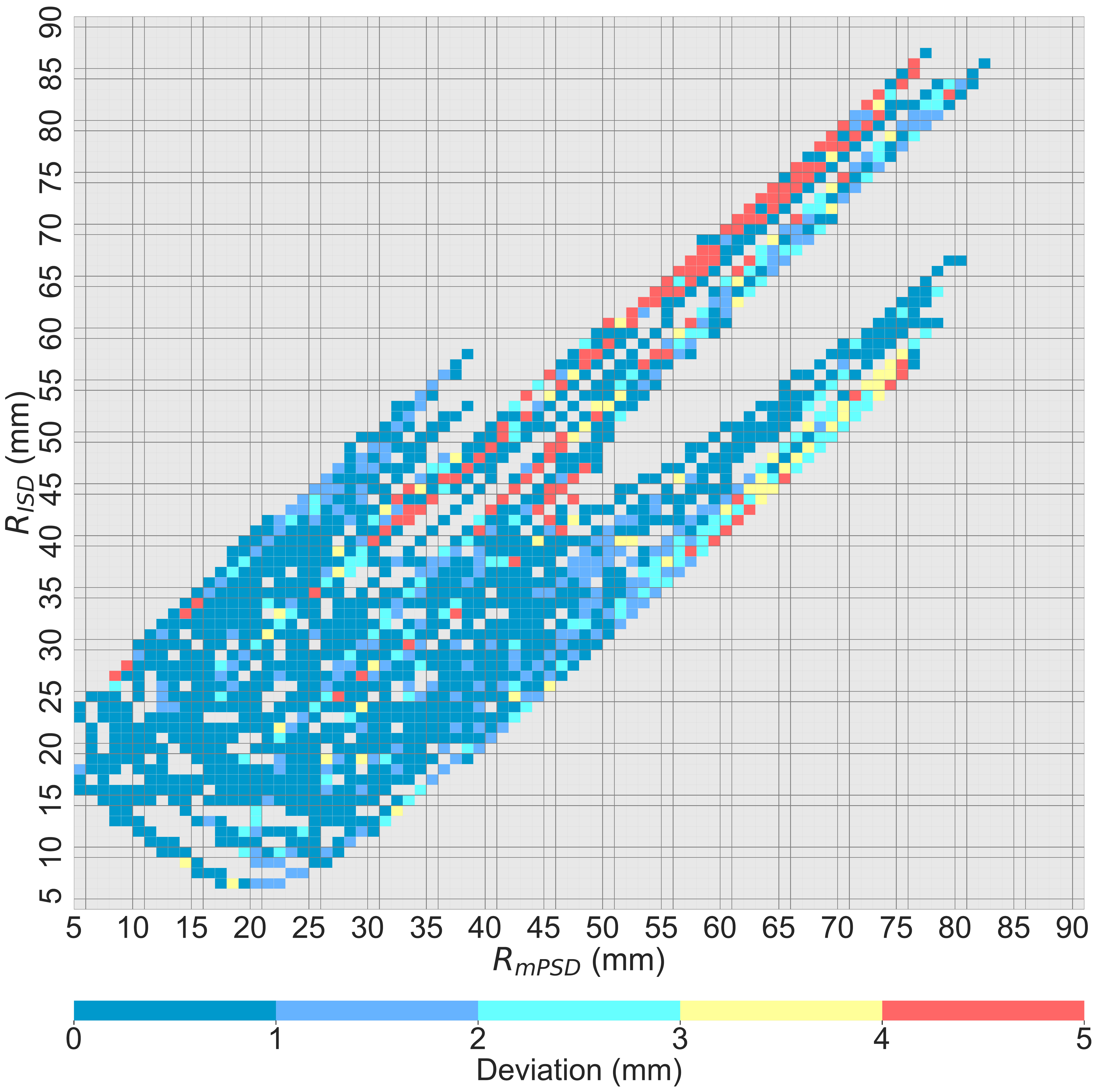}\\
    \end{tabular}
    \caption{\label{R_mPSD_CsI_tracking_heat_map} Source position reconstruction deviations. Heat map of deviations between the measured and expected source locations as a function of the distance to the mPSD-ISD systems. The fields in gray represent the regions where no data is available to quantify the deviations. Please see the online paper version for color's reference.}
    \end{figure}

    The distance dependency of the deviations (2 $\sigma$) shown in in table \ref{table_R_tracking_uncertainty} is given in figure \ref{R_mPSD_CsI_tracking_heat_map}. Figure \ref{R_mPSD_CsI_tracking_heat_map} shows the heat map of the mean deviations between the measured and expected source locations as a function of the distance to the mPSD-ISD systems. The radial distances to the mPSD ($R_{mPSD}$) were taken considering as the origin of the coordinates the position of BCF10.
    
    \subsection{mPSD-ISD positioning in an HDR prostate plan}
    
\begin{figure}[h!]
    \centering
    \begin{tabular}{c}
    \includegraphics[trim = 1mm 1mm 1mm 1mm, clip, scale=0.63]{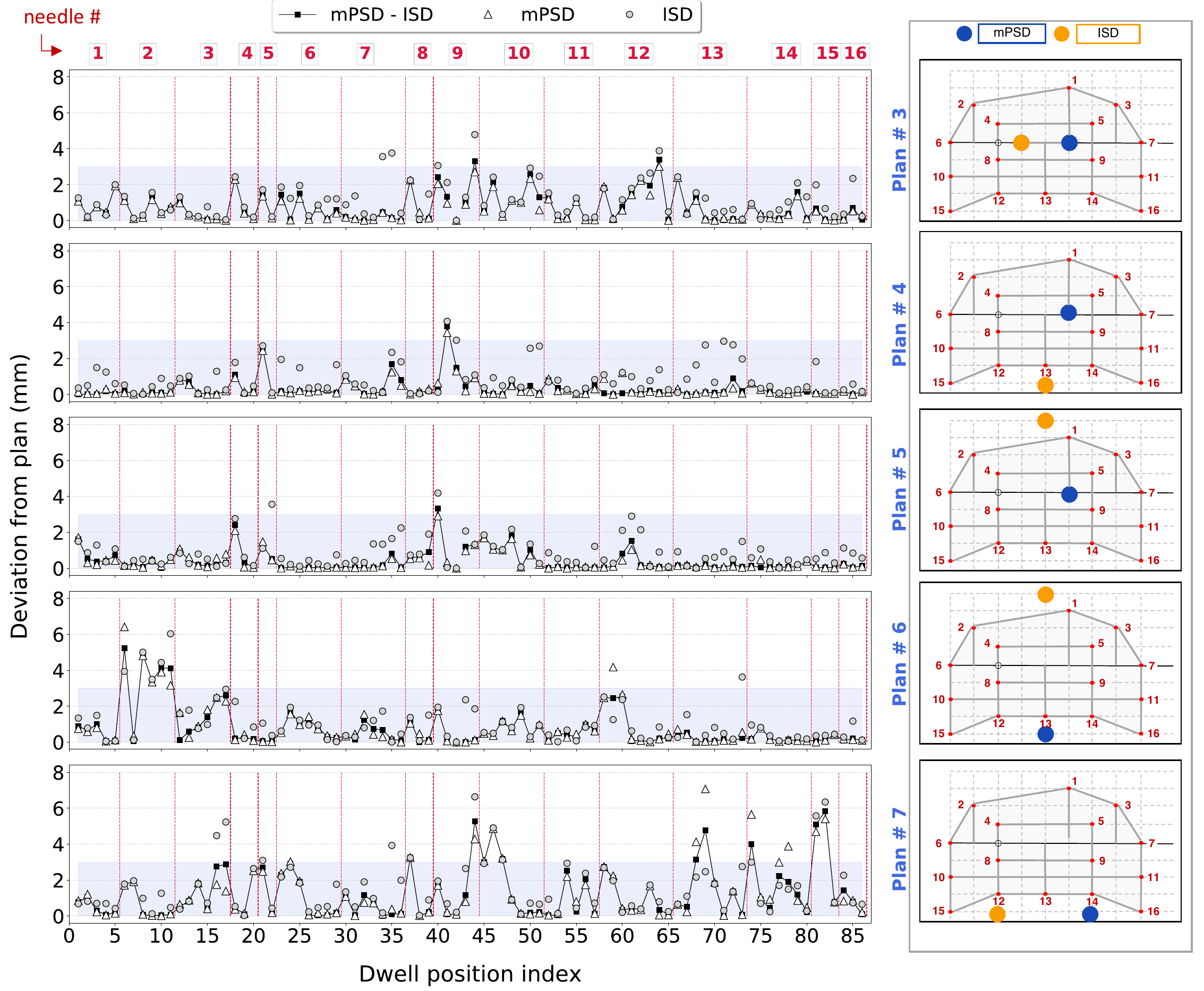}
    \end{tabular}
    \caption{\label{R_mPSD_CsI_prostate_r_deviations} mPSD-ISD deviations from the planned position for the prostate plan (C. \# 3). The squares represent the deviations from the plan when using the combined responses of the mPSD and ISD. The dots represent the independent deviation of the ISD, while the triangles the mPSD's deviations. The vertical lines represent the limits of the data associated to each needle. The annotated values in the graph upper side show the needle numbers. The shaded region represents a 2 mm deviation range. The schematics in the right side of the figure illustrate the plan's numbers as well as the detectors' positions for each configuration.}
    \end{figure}

    Figure \ref{R_mPSD_CsI_prostate_r_deviations} summarizes the results obtained for irradiations with a prostate template (\# 3 in table \ref{table_MM_Irr_plan_info}). The y-axis represents the point-to-point deviations between the mPSD-ISD's reconstructed source location and the planned position. The x-axis represents each dwell position index number. The vertical lines delimit each needle's data. For easy understanding and visualization of the results obtained, the schematics in the right panels of figure \ref{R_mPSD_CsI_prostate_r_deviations} illustrate the plan's number and for each the needle numbers as well as the detectors' positions. The squares in figure \ref{R_mPSD_CsI_prostate_r_deviations} represent the deviations from the plan when using the combined responses of the mPSD and ISD. The dots and triangles represent the independent contribution of the mPSD and ISD to the observed combined deviation. Overall, the accuracy in reconstructing the source position remains within 3 mm from the plan, especially if at least one dosimeter is placed centrally in the prostate. 
    
    \subsection{Detection of positioning errors}
    
\begin{figure}[h!]
    \setlength{\tabcolsep}{1pt}
    \centering
    \begin{tabular}{c}
    \includegraphics[trim = 1mm 1mm 1mm 1mm, clip, scale=0.4]{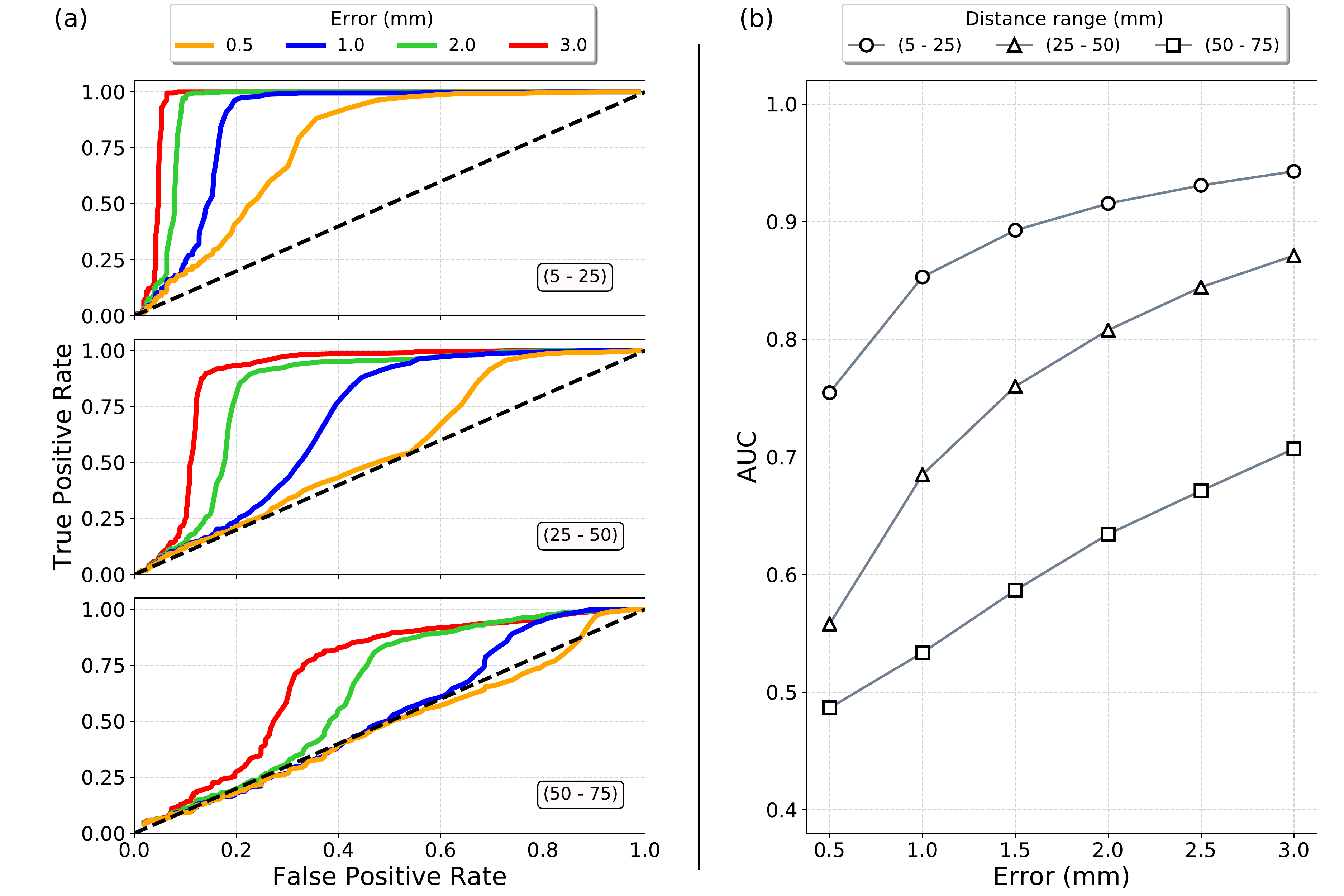}
    \end{tabular}
    \caption{\label{R_mPSD_CsI_ROC} mPSD-ISD error detection probability as a function of the distance to the source. (a) ROC analysis. (b) AUC. The dashed line in (a) shows the limit for random detection probability.}
    \end{figure}

    Figure \ref{R_mPSD_CsI_ROC} summarises the key results obtained when quantifying the error detection probability as a function of the distance to the source. Figure \ref{R_mPSD_CsI_ROC}(a) shows the ROC curves obtained for the different ranges of distances to the source. For visualization purposes, only the curves for introduced error of 0.5, 1.0, 2.0 and 3.0 mm are illustrated. Figure \ref{R_mPSD_CsI_ROC}(b) shows the areas under the curve (AUC) as a function of the full range of introduced errors as well as for the three categories of source-to-dosimeter distances.
    
\begin{figure}[h!]
    \setlength{\tabcolsep}{1pt}
    \centering
    \begin{tabular}{c}
    \includegraphics[trim = 1mm 1mm 1mm 1mm, clip, scale=0.58]{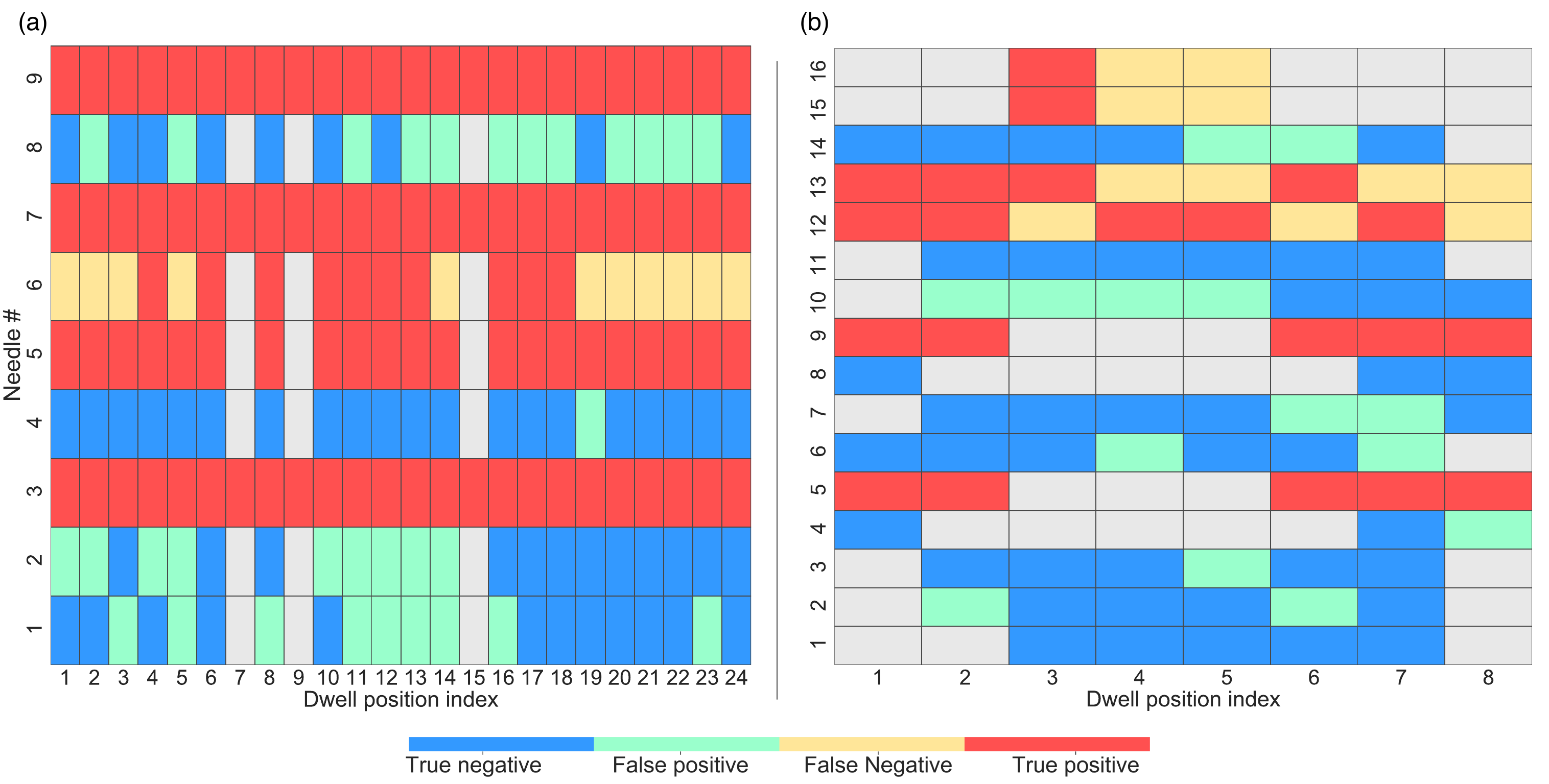}
    \end{tabular}
    \caption{\label{R_Error_heat_map} Heat map of deviations classified into errors for plan number (a) 1 and (b) 3 when using the distance-based threshold to classify deviations as treatment' errors. The fields in gray correspond to not programmed dwell positions.}
    \end{figure}

    Figure \ref{R_Error_heat_map} shows the heat map of deviations translated to an error probability according to the binary classifiers extracted from the contingency table for plans 1 and 3. The x-axis shows the dwell position index and the y-axis the needle number according to the schematic shown in figure \ref{simult_dose_meas_scheme}. The fields in gray in figure \ref{R_Error_heat_map} refer to non-programmed dwell positions.
    
    Table \ref{table_R_Error} summarizes the results of the binary classification done over plan number 1 and 3 for error detection, where we show the results for the distance-based threshold selection as well as the results for fixed error's thresholds used in all the range of distances to the source. Active dwells in table \ref{table_R_Error} refers to the total number of active dwell positions on the plan used as the reference. The sensitivity expresses the proportion of actual positives that are correctly identified, while the specificity, the proportion of actual negatives that are correctly identified. The accuracy reflects the fraction of the measurement in agreement with what was expected. The sensitivity, specificity and accuracy are statistical measures of the performance of a binary test such as the classification of an in vivo measurement into “No alarm” or “Alarm” \cite{Tanderup-invivo-Brachy-2013}.
    
    \begin{table}[h!]
    \caption{\label{table_R_Error} Summary of error detection probabilities found for plan 1 and 3.}
    \setlength{\tabcolsep}{6 pt}
    \centering
    \resizebox{\textwidth}{!}{
    \vspace{0.2cm}
    	\begin{tabular}{c|ccccccccc}
    	\hline
    	\hline
    	\textbf{Acceptance} &\multirow{2}{*}{\textbf{Plan}} & \textbf{Active}  &	\textbf{True}  &  \textbf{False}  &\textbf{True}  &	\textbf{False} & \multirow{2}{*}{\textbf{Sensitivity}}  &	\multirow{2}{*}{\textbf{Specificity}} &	\multirow{2}{*}{\textbf{Accuracy}} \\
    	 \textbf{threshold} & & \textbf{dwells}  &	\textbf{Negatives}   &	\textbf{Positives}   &\textbf{Positives} &	\textbf{Negatives}	&               &               &	         \\
    	 \textbf{(mm)}& & \textbf{(\#)}    &	\textbf{(\#)}        &	\textbf{(\#)}        &\textbf{(\#)}      &	\textbf{(\#)}    	& \textbf{(\%)}          &	\textbf{(\%)}        &	\textbf{(\%)}      \\
        \hline
        1 
        & \multirow{4}{*}{\textbf{1}} 
        & \multirow{4}{*}{\textbf{189}}    
        & 23 
        & 61 
        & 114 
        & 0  
        & 100.00	
        & 27.38  
        & 69.19   \\
        
        2 
        & & & 51
        & 33 
        & 114 
        & 0 
        & 100.00	
        & 60.71  
        &   83.33   \\
        3
        & & & 68 
        & 16 
        & 103 
        & 11 
        & 90.35	    
        & 80.95  
        & 86.36   \\
        
        1, 2, 3 $^{*}$ 
        & & & 53 
        & 31 
        & 103 
        & 11 
        & 90.35	    
        & 63.10  
        & 78.79   \\
        \hline 
        1 
        &     \multirow{4}{*}{\textbf{3}} 
        & \multirow{4}{*}{\textbf{86}}    
        & 16 
        &	41 
        &	30 
        &	2  
        &	93.75     
        & 28.07  
        &	51.69   \\
        
        2 
        & & & 36 
        &	21 
        &	24 
        &	8  
        &	75.00     
        & 63.16  
        &	67.42   \\
        
        3 
        & & & 45 
        &	12 
        &	14 
        &	18 
        &	43.75     
        & 78.95  
        &	66.29   \\
        
        1, 2, 3 $^{*}$ 
        & & & 43 
        &	14 
        &	21 
        &	11 
        &	65.62     
        & 75.44  
        &	71.91   \\
        \hline
        \multicolumn{10}{l}{\scriptsize{$*$ Distance-based thresholds selection: 1 mm for dwells within 5-25 mm range, 2 mm for dwells within 25-50 mm range and 3 mm for dwells beyond 50 mm range.}}
    	\end{tabular}}
    \end{table}

\section{Discussion}

\subsection{Measurement uncertainty influence on the source position reconstruction accuracy}

    The inclusion of the measurement uncertainty in the source location calculation leads to the improvement on the source location estimation accuracy. In such a case the highest densities and median values were found close to 0 (values close to 0 and a small dispersion around it represent better agreement with the reference). When the uncertainty was excluded from the tracking process, no intersections were identified between the mPSD and ISD based distances in 22.8  \% of the cases. In contrast, intersections were always identified when the measurement uncertainty was considered. However, only 73(96) \% of the expected dwell positions were found within the intersection band for 1(2) $\sigma$ uncertainties. Thus, considering a 2$\sigma$ criteria increases the overlap regions between the two dosimeters and leads to a better agreement with the expected position in all axes. 
    
\subsection{Source position reconstruction accuracy as function of the  distance to the source}
\label{D_source_pos_reconst}

    In this study, we covered up to 12 cm range of source movement around the mPSD-ISD detectors. Figure \ref{R_mPSD_CsI_tracking_heat_map} provides an idea on the experimental limits of the source tracking with the mPSD-ISD systems. For distances from the source grater than 5 cm, measurement uncertainties and possibly inaccuracies in the energy correction for the inorganic crystal lead to large positional uncertainties, between 3 and 5 mm from the planned position. For most clinical cases, dwell positions beyond 5-6 cm are rarely observed. However, finding HDR treatments with the $^{192}Ir$ sources stepping within a range of 5-6 cm is common in a clinical context, especially for prostate cases.
    
    Several studies \cite{Fonseca-2017, Cartwright-2010, Hardcastle-MOSkin-2010, Kertzscher-2011, Seymour-2011, Kertzscher-2014, Johansen-2018, Sethi-Doppler-US-2018, Guiral-2016, Smith-invivo_Brachy_EPID-2013} have focused on the source tracking topic in HDR BT.  The common point between those studies is the degeneracy in one of the detector axes, leading to source tracking report in a plane, not a point in the space. Some other authors have proposed different methods based on pinholes collimation panels to find the source location in 3D coordinates \cite{Duan-PinholeCamera-tracking-2001, Bativc-2010, Safavi-BrachyView-2015,Watanabe-pinhole-tracking-2018}. This is the first study to report on 3D source tracking based on point detector dosimetry. By combining the mPSD and ISD responses, we broke that degeneracy, enabling a determination the source location. The method here proposed can be extended to the combination of different systems. There is no restriction about which detector should be used for such a set-up.

\subsection{mPSD-ISD positioning in an HDR prostate plan}
    
    The reconstruction of the source position in the prostate template (configuration \# 3 in table \ref{table_MM_Irr_plan_info}) showed an overall agreement with the planned position of 3 mm or better for most dwell positions. Having at least one of the two detectors located inside the the prostate volume (plan \# 3, 4, 5), reduces the measurement's deviations considerably. However, an increased deviation is observed in the experiments done with at least one of the detectors at distal positions from the prostate volume. As could be observed in plan \# 6 and \# 7 in Fig. \ref{R_mPSD_CsI_prostate_r_deviations}, deviations up to 7 mm were seen, with the worst results in plan \# 7, where both dosimeters are positioned out of the prostate volume. Thus selecting the optimal location for the detector's positioning within a clinical volume needs to be based on the system uncertainty and in particular the overall performance as a function of distances from the source.

\subsection{Detection of positioning errors}
    
     In agreement with the study reported by Andersen et \textit{al.} \cite{Andersen-time-resolved-2009}, as the detector-to-source distance increases, the capability to detect errors is affected. Because of an increase of the measurement uncertainties, at long distances to the source, large errors should occur, to be detected by the dosimetry systems. This effect is shown in figure \ref{R_mPSD_CsI_ROC}. At short distances, there is a high probability of detecting positioning errors of 1 mm or more, with a low incidence of false alarms. As could be observed in figure \ref{R_mPSD_CsI_ROC}(a), for the dwell positions with distances to the source beyond 50 mm, more than 3 mm of deviations has to be accepted if considering the FPR-TPR criterion previously set. However, according to the recommendations established for HDR BT practices \cite{Fonseca-IVD-Brachy-2020}, we decided to set the threshold to 3 mm of detected deviation relative to the planned position. The AUC was found higher than 0.85 and 0.68 at ranges of 5 - 25 and 25-50 mm from the source, respectively. However, for distances to the source within 50 - 70 mm range, the detection of 0.5 and 1.0 mm errors was essentially consistent with random results (AUC around 0.5), and only large error of 3 mm or more can be detected with some confidence (i.e. reasonable the false alarm rates).
     
     Figure \ref{R_Error_heat_map} and table \ref{table_R_Error} summarize our findings when evaluating the detection of positioning errors and the \textit{distance-based threshold selection} in plan number 1, and 3. mPSD-ISD's best performance was observed in plan \# 1. The accuracy observed was 78.79\% compared to 71.91\% for plan \# 3. The sensitivity in plan \# 1 is also superior. Several factors can explain such behaviour. First of all, from the schematics in figure \ref{simult_dose_meas_scheme}, we can infer that the needles distribution is more complex in configuration \# 3 than configuration \# 1. In the latter, a single intersection band (see figure \ref{intersection_scheme_CsI_mPSD}c for reference) of its responses was found in 92 \% of the dwell positions. In configuration \# 3, the source position reconstruction was complicated by the symmetrical arrangement of needles  relative to the mPSD and ISD (needles 4, 5, 8 and 9). In most of the dwell positions in needle 4, 5, 8 and 9, we found a dual-band behaviour. Thus the swap in those pairs of needles becomes difficult to detect if there is, for instance, similar ``needles identities" (\textit{i.e} having the same dwell positions planned). Thus, the second factor influencing the error detection capability is the needles arrangement and dwell position distribution around the mPSD-ISD locations. Thus detector positioning appears an important factor influencing outcome. In needle 5 and 9 in figure \ref{R_Error_heat_map}b, we were able to correctly detect the needle swap because, in needle 5, only the position 1 and 2 were active, while in needle 9, the active dwell positions were 6 to 8. Overall, dosimeter configurations shown in plan \# 4 and \# 5 constitute better choices. Based on these results, we recommend to have a dosimeter close to the central portion of the prostate and one in or very close to an OAR, for instance rectum or urethra. Fig. 7 shows the importance of having a dosimeter placed centrally, and our ROC analysis shows that a lower threshold can be set close to the dosimeter. Hence placing a dosimeter in OAR could put priority to that OAR.
     
     Finally, the impact of using different thresholds (1,2 and 3 mm) for considering a deviation in the source position reconstruction as an error has been investigated (e.g. table \ref{table_R_Error}). The use of 1 mm thresholds for all source-dosimeter distances improves the sensitivity. However, it compromises the specificity. For any system, there is a trade-off between sensitivity and specificity since the number of false alarms is increased if the sensitivity is increased by lowering the threshold for error detection \cite{Tanderup-invivo-Brachy-2013}. The best balance sensitivity/specificity/accuracy in plan number 1 is obtained for a deviation threshold of 3 mm, while in plan number 3 is obtained with the distance-based threshold selection. The advantage of a distance-based threshold is that it leverage the maximum performance (as underlined by the AUC results) of a given dosimeter, taking into account its associated uncertainty chain. In a clinical setting, thresholds should be chosen to allow potentially clinically significant errors to be detected while accepting that this will result in some unnecessary treatment interruptions or post-treatment investigations.
     
     In this paper, we solely used the threshold values extracted from the ROC analysis to classify deviations as errors. However, considering only the threshold criterion in combination with measurement uncertainty led to the miss-classification of an event into false positives/negatives. In this sense, if using, for instance, the distance-based threshold method, we found that from the 67 dwell positions classified as false positives/negatives in plan number 1 and 3, the difference between reconstructed source position deviation from the plan and the classification threshold was smaller than 0.5 mm for 35 of them. Trusting the results for a single dwell position will probably represent an unnecessary treatment interruption. A more robust algorithm for error classification has to be considered to minimize unnecessary treatment interruptions without forgetting the clinical significance of deviations. A solution to reduce the number of false alarms could be the inclusion of the source dwell time measurements in combination with the thresholds approach on source position deviations.
     
\section{Conclusions}
    
    In this study, we performed a 3D source position reconstruction by combining in vivo dosimetry measurements from two independent detector systems. By combining the mPSD and ISD responses, we broke the degeneracy in the detectors' radial direction, enabling a determination of the source location. The inclusion of the measurement uncertainty in the source location calculation leads to improved source location estimation accuracy. For a range of distances within 50 mm from each system, the deviations in the source position reconstruction are below 3 mm from the planned position. The measurements in an HDR prostate plan configuration demonstrated the effectiveness mPSD-ISD combination for measurements in a range of distances to the source with clinical relevance. We recommend the positioning of one detector in the center of the prostate volume and the other outside of it. We also recommend a non-symmetrical distribution of the treatment needles around the detectors, which would allow the real-time detection of positioning errors with a high rate of accurate classification, while keeping the false alarms rate low.
    
\section{Acknowledgement}
    The present work was supported by the National Sciences and Engineering Research Council of Canada (NSERC) via the NSERC-Elekta Industrial Research Chair grants Nos. 484144-15 and RGPIN-2019-05038, and by a Canadian Foundation for Innovation (CFI) JR Evans Leader Funds grant \# 35633. Haydee Maria Linares Rosales further acknowledges support from Fonds de Recherche du Quebec - Nature et Technologies (FRQ-NT) and by the CREATE Medical Physics Research Training Network grant of the Natural Sciences and Engineering Research Council of Canada (Grant \# 432290).
      
\cleardoublepage

\bibliographystyle{ieeetr}

\cleardoublepage
\end{document}